\documentclass[conference]{IEEEtran}

\usepackage{amsmath,amssymb,amsfonts}
\usepackage{algorithmic}
\usepackage{graphicx}
\usepackage{textcomp}
\usepackage{xcolor}
\usepackage{multirow}
\usepackage[section]{placeins}

\AtBeginDocument{%
  }

\begin{document}

\title{4D Multimodal Co-attention Fusion Network with Latent Contrastive Alignment for Alzheimer's Diagnosis}

\author{Yuxiang Wei, Yanteng Zhang, Xi Xiao, Tianyang Wang, Xiao Wang, Vince D. Calhoun}
\maketitle


\begin{abstract}
Multimodal neuroimaging provides complementary structural and functional insights into both human brain organization and disease-related dynamics. Recent studies demonstrate enhanced diagnostic sensitivity for Alzheimer's disease (AD) through synergistic integration of neuroimaging data (e.g., sMRI, fMRI) with behavioral cognitive scores tabular data biomarkers. However, the intrinsic heterogeneity across modalities (e.g., 4D spatiotemporal fMRI dynamics vs. 3D anatomical sMRI structure) presents critical challenges for discriminative feature fusion. To bridge this gap, we propose M2M-AlignNet: a geometry-aware multimodal co-attention network with latent alignment for early AD diagnosis using sMRI and fMRI.  At the core of our approach is a multi-patch-to-multi-patch (M2M) contrastive loss function that quantifies and reduces representational discrepancies via geometry-weighted patch correspondence, explicitly aligning fMRI components across brain regions with their sMRI structural substrates without one-to-one constraints. Additionally, we propose a latent-as-query co-attention module to autonomously discover fusion patterns, circumventing modality prioritization biases while minimizing feature redundancy. We conduct extensive experiments to confirm the effectiveness of our method and highlight the correspondance between fMRI and sMRI as AD biomarkers.
\end{abstract}

\maketitle

\section{Introduction}
Alzheimer's disease (AD) represents a progressive neurodegenerative disorder where pathological changes precede clinical manifestations, making early detection crucial for therapeutic intervention \cite{deture2019neuropathological}. The pre-symptomatic stage of AD, characterized by the presence of disease biomarkers without cognitive impairment, has become a focal point of research. Key pathological biomarkers, such as amyloid-$\beta$ peptide ($A\beta_{42}$), amyloid-PET (AV45), total Tau (tTau), and phosphorylated species of Tau (pTau) have been identified as critical risk factors for pre-symptomatic AD \cite{tandon2023predictors}. 

Functional magnetic resonance imaging (fMRI) and structural MRI (sMRI) serve as pivotal non-invasive tools for monitoring AD progression, offering distinct yet complementary insights into brain structure and function \cite{qiu20243d}. Advanced deep learning architectures, particularly 3D CNNs \cite{tan2019efficientnet} and Transformer \cite{kim2023swift} have demonstrated remarkable capabilities in analyzing these neuroimaging data. However, reliance on a single imaging modality constrains models’ ability to capture complementary structural and functional perspectives. Consequently, recent studies prioritize multimodal fusion methods to enable holistic understanding of brain patterns. While existing studies successfully combine spatial modalities such as sMRI+PET or temporal modalities such has fMRI+EEG \cite{qiu20243d,abrol2019multimodal,rahim2023prediction,ning2021relation,liu2024mcan}, few method adequately addresses the fundamental challenge of integrating 4D spatiotemporal fMRI (dynamic activity) with 3D anatomical sMRI (static structure). 

A common fusion approach involves direct addition or concatenation, applied either in the input space or latent feature space. However, this strategy overlooks the intrinsic heterogeneity of brain modalities. For example, fMRI captures 4D dynamic brain activity by measuring blood oxygenation-dependent signal while sMRI captures 3D information about tissue types and anatomical structures. The misalignment could result in suboptimal fusion.

Beyond direct fusion, prior studies explore co-attention to guide the encoding of individual modality within a \textit{modality-as-query} paradigm \cite{liu2024mcan,ding2024multimodal}, where one modality dictates cross-modal interactions (e.g., using sMRI embeddings as queries to attend to PET features). Nevertheless, this rigid modality prioritization could contradict the inherent complementary nature between multiple modalities. Similarly, contrastive learning \cite{radford2021learning} enforces intra-subject matching through pairwise alignment of multimodal instances. However, this fails to capture the brain's distributed many-to-many interdependencies, where functional dynamics emerge from non-linear interactions between multiple structural substrates. Clinical evidence further corroborates these complex relationships between functional networks and structural components in the brain \cite{khalilullah2023multimodal}, underscoring the inadequacy of rigid hierarchies or simplistic contrastive pairs.  

To address these gaps, we propose \textbf{M2M-AlignNet}: a novel multimodal co-attention fusion network with geometry-aware multi-patch-to-multi-patch (M2M) latent alignment. First, our geometry-weighted alignment module maximizes cross-modal patch similarity using distance-adjusted correspondences, explicitly modeling fMRI-sMRI interactions beyond one-to-one constraints. Next, a latent-as-query co-attention mechanism dynamically fuses decomposed fMRI features with sMRI and tabular data, eliminating assumptions about modality hierarchies. Finally, a bottleneck module selectively condenses cross-modal features while suppressing redundancy. Through extensive experiments, we demonstrate superior performance over baselines, supported by ablation studies and interpretability analysis revealing fMRI-sMRI correspondences. Our contributions are threefold:
\begin{itemize}
    \item Spatiotemporal fusion framework: We pioneer fusion of 4D fMRI and 3D sMRI, bridging dynamic processes with structural foundations.
    \item M2M contrastive alignment: We systematically model many-to-many functional-structural relationships via geometry-weighted inter-subject alignment. 
    \item Latent-as-query co-attention: We use trainable latent queries to autonomously discover fusion patterns, circumventing modality prioritization biases while minimizing feature redundancy. 
\end{itemize}

\section{Related Works}
\subsection{Multimodal Fusion of Brain Images}
The integration of multiple imaging modalities offers complementary insights, enhancing the understanding of human cognitive functions and neuro-disorders. Most existing studies focus on fusing spatial information from imaging modalities such as sMRI and positron emission tomography (PET). For example, Qiu et al. \cite{qiu20243d} utilized 3D ResNet to extract features from PET and sMRI, followed by dedicated modules for local and global modality fusion. Similarly, Ning et al. \cite{ning2021relation} projected sMRI and PET into a shared latent space and calculated a reconstruction loss for training. A sample can be directly predicted as AD by applying a projection matrix over the latent representations. For temporal fusion, Liu et al. \cite{liu2024mcan} explored the integration of time-series data by designing an attention-based encoder to fuse functional MRI (fMRI) with EEG at an early stage, subsequently combining predictions from individual modalities and their shared representation.

While these studies primarily focus on either spatial or temporal fusion, limited research has addressed the integration of 4D spatiotemporal and 3D spatial modalities, such as fMRI and sMRI. In addition, early multimodal approaches often assumed equal contributions from different modalities, using simple addition or concatenation for fusion at various network stages \cite{abrol2019multimodal}. Although recent works employ co-attention to model pair-wise correlations between modalities \cite{liu2024mcan,ding2024multimodal}, the misalignment of modalities in the latent space remains a challenge due to their inherent heterogeneity. In this work, we propose a patch-wise contrastive loss to explicitly align fMRI with sMRI in the latent space. This alignment fosters homogeneity in fused representations, enabling more effective integration of spatiotemporal and spatial information for downstream tasks.

\subsection{Contrastive Learning}
Contrastive learning has been widely applied in self-supervised learning and multimodal representation learning due to its ability to learn meaningful embeddings by contrasting positive and negative pairs. A representative loss function is InfoNCE \cite{oord2018representation}, which uses an entropy-based divergence that effectively pulls positive pairs closer in the embedding space while pushing negative pairs apart. In the multimodal setting, CLIP \cite{radford2021learning} represents a prominent approach that aligns image and text embeddings by maximizing their pairwise cosine similarity. It employs a symmetric cross-entropy loss to match an image with its corresponding text description, thereby learning a shared representation space for both modalities. 

Inspired by the previous works, we introduce a multi-patch-to-multi-patch (M2M) contrastive loss to align fMRI with sMRI in the latent space. Different from \cite{radford2021learning} that enforces intra-subject matching through pairise alignment of multimodal instances, our approach performs patch-wise alignment across subjects, pulling corresponding embedding patches closer while pushing irrelevant patch pairs apart. Given the unknown correlations between fMRI and sMRI structures, our method incorporates an adaptive alignment mechanism. This mechanism allows multiple fMRI patches to align with multiple sMRI patches dynamically, leveraging a discrepancy-based self-weighting strategy to assign importance to each patch pair during alignment. This adaptive weighting ensures that structurally significant patches contribute more heavily to the alignment process, enhancing the robustness of the learned representations.

\section{Methodology}
\begin{figure*}[ht]
\centerline{\includegraphics[width=\textwidth]{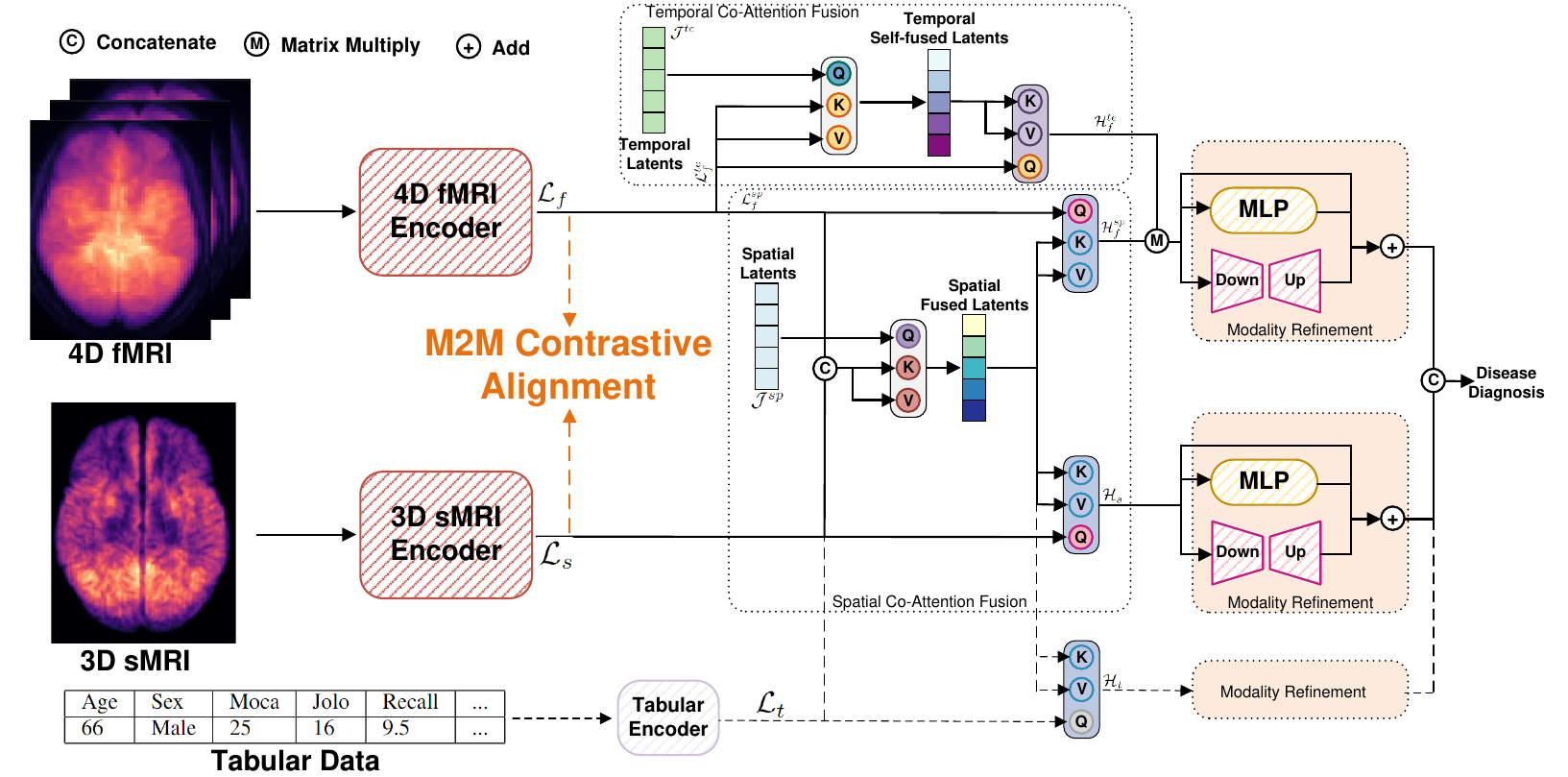}}
\caption{M2M-AlignNet: Modalities are first encoded by the corresponding modality-specific encoders, then fused via co-attention. fMRI and sMRI representations are further aligned in the latent space via M2M contrastive loss.}
\label{fig:framework}
\end{figure*}

The overall framework of the proposed method is shown in Fig. \ref{fig:framework} and can be summarized as follows: 1) 4D Swin Transformer-based backbone to encode fMRI and sMRI into latent space. 2) M2M contrastive alignment loss to align fMRI with sMRI in latent space. 3) Spatial and temporal co-attention fusion with bottlenecks to fuse modalities. Further details are provided in the following sections.

\subsection{Materials and Proprocessing}
In this work, we conduct experiments based on the Emory Healthy Brain Study (EHBS) dataset \cite{goetz2019rationale}, which is a longitudinal cohort study of cognitively normal adults (50-75 years) with a risk of developing AD. A total of 642 subjects are selected with available fMRI scans, with 545 biomarker-negative (CN) and 97 biomarker-positive (pre-symptomatic) subjects. We use tTau and $\mathrm{A\beta_{42}}$ as biomarkers to identify Psym from CN, where subjects with $\frac{\mathrm{tTau}}{\mathrm{A\beta_{42}}} > 0.24$ are identified as at-risk for AD \cite{tandon2023predictors}.

The fMRI and sMRI data underwent preprocessing with SPM12, which involved correcting head movement, aligning time, aligning data to the Montreal Neurological Institute (MNI) space using an echo planar imaging (EPI) template, resampling to 3 mm isotropic voxels, and applying 6 mm full width at half maximum Gaussian smoothing. 

\subsection{Modality-Specific Feature Extraction}
We choose SwiFT \cite{kim2023swift}, a state-of-the-art 4D Swin Transformer model as the backbone to encode MRI features into latent space. SwiFT is a hierarchical model that gradually extracts features from high-dimensional fMRI using self-attention. Since it is designed to process fMRI, we add a temporal dimension to sMRI and employ the SwiFT with the same parameters to encode the two modalities. The number of channels in each stage is set to [24, 48, 96, 96]. Finally, we obtain the latent features $\mathcal{L}_f,\mathcal{L}_s\in \mathbb{R}^{c\times h \times w \times d \times t}$, where $t=1$ for sMRI.

For the tabular data, we apply a simple multi-layer perception (MLP) that encodes the features to have the same dimension as MRI features, $\mathcal{L}_t \in \mathbb{R}^{c \times l}$.

\subsection{Spatial and Temporal Co-Attention Fusion}
To effectively fuse modalities and utilize the complementary information, inspired by \cite{liu2024mcan}, we propose a co-attention-based fusion module with learnable embeddings. For fMRI, since $\mathcal{L}_f$ is a 4D embedding with a temporal dimension, we first decompose its spatial and temporal components by averaging the corresponding dimension, which results in $\mathcal{L}_f^{sp} \in \mathbb{R}^{c \times h \times w \times d}$ and $\mathcal{L}_f^{te} \in \mathbb{R}^{c \times t}$.

To fuse modalities, the fMRI, sMRI, and tabular embeddings are first flattened and concatenated as the key and value, then the co-attention is computed with an additional learnable latent query $\mathcal{J}^{sp}$:
\begin{equation}
    \mathcal{H}_{joint}^{sp} = \mathrm{softmax}(\frac{W_Q^{sp}\mathcal{J}^{sp} \cdot (W_K^{sp}\mathcal{L}_{joint}^{sp})^T}{\sqrt{c}}) \cdot W_V^{sp} \mathcal{L}_{joint}^{sp}
\label{eq:lFuse}\end{equation}
where $\mathcal{L}_{joint}^{sp}$ is the concatenation of modality representations, $W_{Q,K,V}^{sp}$ are the weight matrices. Note that $\mathcal{J}^{sp}$ initialized as a normal distribution with the dimension as $\mathcal{L}_{joint}^{sp}$.

The fused latent representation $\mathcal{H}_{joint}^{sp}$ is used as a guidance to refine modality-specific information, allowing the model to share complementary information across modalities while preserving individual modality's unique properties:
\begin{equation}
\begin{aligned}
    &\mathcal{H}_{f}^{sp} = \mathrm{softmax}(\frac{W_{Q,f}^{sp} \mathcal{L}_f^{sp} \cdot (W_{K,f}^{sp}\mathcal{H}_{joint}^{sp})^T}{\sqrt{c}}) \cdot W_{V,f}^{sp} \mathcal{H}_{joint}^{sp} \\
    &\mathcal{H}_{(s,t)} = \mathrm{softmax}(\frac{W_{Q,(s,t)}^{sp} \mathcal{L}_{(s,t)} \cdot (W_{K,(s,t)}^{sp}\mathcal{H}_{joint}^{sp})^T}{\sqrt{c}}) \\
    &\quad \quad \quad \quad \quad \quad \quad \quad \quad \quad \quad \quad \quad \quad \quad \quad \quad  \cdot W_{V,(s,t)}^{sp} \mathcal{H}_{joint}^{sp}
\end{aligned}
\label{eq:sFuse}\end{equation}

Since only fMRI contains temporal information, we propose to refine the features with a similar latent co-attention "self-fusion" module. Specifically, $\mathcal{L}_f^{te}$ is first fused with a temporal latent embedding $\mathcal{J}^{te}$ as in Eq. \ref{eq:lFuse} and produce $ \mathcal{H}_{f}^{te}$. Subsequently, similar to \ref{eq:sFuse}, "self-fusion" is performed to refine temporal information. The resultant representation $\mathcal{H}_{f}^{te}$ is multiplied with $\mathcal{H}_{f}^{sp}$ to combine space and time and generate the spatiotemporal embedding.

Because the information from multiple modalities can be redundant, we employ a modality refinement module to condense fused features from each modality \cite{nagrani2021attention}. As in Fig. \ref{fig:framework}, the module contains a simple one-layer MLP, a down-up bottleneck, and a residual connection. The down-up bottleneck is implemented with simple linear projections $P$ with GELU activation function to promote non-linearity:
\begin{equation}
    \mathrm{Bottleneck}(\mathcal{H}_{(f,s,t)})= P_{up}(\mathrm{GELU}(P_{down}(\mathcal{H}_{(f,s,t)})))
\end{equation}
where $P_{down} \in \mathbb{R}^{c \times \frac{c}{4}}$ and $P_{up} \in \mathbb{R}^{\frac{c}{4} \times c}$.

The refined representations from all modalities are concatenated and projected to diagnose AD.

\begin{figure}[t]
\centerline{\includegraphics[width=\columnwidth]{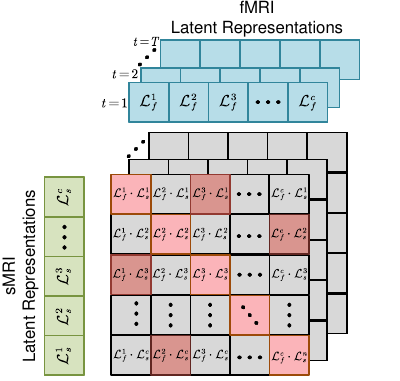}}
\caption{M2M contrastive loss to align pairs of fMRI patches with sMRI patches at each time point. Multiple fMRI patches can be aligned with multiple sMRI patches.}
\label{fig:cont}
\end{figure}

\subsection{M2M Contrastive Alignment}
Due to the inherent differences between fMRI and sMRI, we propose to explicitly align them in the latent space. Previous research, while primarily focusing on aligning samples from one modality with samples from another \cite{radford2021learning,lyu2024omnibind,zhu2023languagebind}, showcasing the potential of modality alignment in uncovering cross-modality relationships and generating a coherent understanding of the data. Instead of performing batch-level alignment, we perform patch-level level alignment to facilitate the fusion of fMRI and sMRI.

Fig. \ref{fig:cont} illustrates a summary of M2M contrastive loss. After encoding fMRI and sMRI data into the latent space, at each time point $t$ of fMRI, we flatten the representations and compute the paired patch-wise similarity between each patch from $\mathcal{L}_f$ with each patch from $\mathcal{L}_s$:
\begin{equation}
    S^t \in \mathbb{R}^{C \times C}, S^{t,(i,j)} = s(\mathcal{L}_f^{t,(i)}, \mathcal{L}_s^j) = \mathcal{L}_f^{t,(i)} (\mathcal{L}_s^j)^T
\end{equation}

Inspired by contrastive learning \cite{chen2020simple}, we aim to maximize the agreement between corresponding positive pairs while enlarging the discrepancy between negative pairs. The loss for a positive patch pair $(i,j)$ can be defined as:
\begin{equation}
    l^{t,(i,j)}=-\mathrm{log}\frac{\mathrm{exp}(S^{t,(i,j)} / \tau)}{\sum_{k=1}^{C} \mathbb{1}_{k \neq i}\mathrm{exp}(S^{t,(i,k)} / \tau)}
\end{equation}
where $\mathbb{1}_{k \neq i} \in \{0,1\}$ and equals 1 iff $k \neq i$, $\tau$ is a temperature hyperparameter that controls the distribution. The final loss is computed across all positive pairs and summed over all time steps, thereby enabling dynamic modeling of cross-modal correspondence.

To determine a positive pair $(i,j)$, a straightforward choice is to set all pairs on the diagonal, i.e., $i=j$, as positive pairs \cite{radford2021learning}, which generates 1 positive pair and $C-1$ negative pairs for each row in $S^t$. This assumes a one-to-one correspondence in the latent space and forces the modality encoders to encode potentially correlated fMRI and sMRI features into a single pair. However, since there is no ground truth for the alignment and previous clinical research suggested a multi-to-multi correspondence between fMRI and sMRI \cite{khalilullah2023multimodal}, we loosed the constraint by introducing the adaptive self-weighting for the negative pairs. Concretely, if two patches from a negative pair contain similar semantics, the contrast between them is adaptively weakened. This allows correspondence between a single patch from the fMRI embedding and multiple patches from the sMRI embedding, and vice versa. As a result, the multi-patch-to-multi-patch contrastive loss is defined as:
\begin{equation}
     l_{M2M}^{t,(i,j)}=-\mathrm{log}\frac{\mathrm{exp}(S^{t,(i,j)} / \tau)}{\sum_{k=1}^{C} w^{t,(i,k)} \mathbb{1}_{k \neq i} \mathrm{exp}(S^{t,(i,k)} / \tau)}
\label{eq:m2m}\end{equation}
The pair-wise weight is updated by:
\begin{equation}
    w^{t,(i,k)} = \mathcal{T}(\mathcal{D}(\mathcal{L}_f^{t,(i)}, \mathcal{L}_s^k))
\end{equation}
where $\mathcal{T}$ is a negative correlation function and $\mathcal{D}$ is a divergence function that measures the discrepancy. Numerous methods can be used to determine the discrepancy $\mathcal{D}(\mathcal{L}_f^{t,(i)}, \mathcal{L}_s^k))$. We experiment with a range of commonly used similarity measurements and divergences, including dot product, cosine similarity, KL divergence, Jensen-Shannon divergence (JSD), and maximum mean discrepancy (MMD). The results are presented in the experimental section.

It should be pointed out that our method can be easily extended to handle more than 2 modalities. For each modality pair, Eq. \ref{eq:m2m} is calculated and all combinations are summed. An extra weight may be leveraged to measure the contribution of alignment in each pair. We leave this part as a future direction.

\section{Exepriments}
\subsection{Experimental Settings}
For all training in this study, we use the AdamW optimizer with the cosine annealing learning rate scheduler, with a learning rate of 0.001. We apply 20 warm-up epochs and set the training epochs to 100. Due to the imbalance of the EHBS dataset, we include two area-under-curve metrics, ROC-AUC and PR-AUC, as well as accuracy, to measure the performances. Besides EHBS, we also include the ADNI dataset \cite{jack2008alzheimer}, which contains 474 subjects for AD vs. healthy control classification, and the Human Connectome Project (HCP) S1200 dataset \cite{van2013wu}, which contains 833 healthy young subjects for sex classification. All fMRI and sMRI images in the datasets undergo the standard preprocessing pipelines, including normalization, registration, and smoothing. For all evaluation, we perform five-fold cross-validation and report the averaged metrics as well as the standard deviation.

\begin{table*}[h]
\caption{Compare with Other Multimodal Methods}
\centering
\resizebox{\textwidth}{!}{
\begin{tabular}{c|ccc|ccc|ccc}
 & & EHBS& & & ADNI& & & HCP&\\
\hline
         & PR-AUC     & ROC-AUC      & Accuracy & PR-AUC     & ROC-AUC      & Accuracy & PR-AUC     & ROC-AUC      &Accuracy \\ \hline
SwiFT-EF & 58.03$\pm$3.1 & 64.13$\pm$3.9 & 79.74$\pm$7.8   & 65.54$\pm$2.8& 64.86$\pm$2.9& 67.31$\pm$4.2& 93.10$\pm$1.9& 93.64$\pm$1.1&87.05$\pm$1.5\\ 
SwiFT-LF & 62.93$\pm$5.6 & 59.28$\pm$3.2 &   69.16$\pm$8.4   & 72.47$\pm$3.1& 73.10$\pm$2.2& 68.88$\pm$3.6& 94.45$\pm$1.8& 95.03$\pm$1.6&87.50$\pm$1.6\\ 
EMV-Net  & 61.49$\pm$4.5 & 60.01$\pm$3.3 &  77.14$\pm$15.2   & 69.19$\pm$3.5& 65.16$\pm$2.5& 60.01$\pm$2.9& 90.10$\pm$2.1& 89.91$\pm$1.8&89.16$\pm$2.2\\ 
mmFormer & 64.06$\pm$5.8  & 59.47$\pm$2.8 &   74.30$\pm$16.9      & 75.02$\pm$3.1& 75.08$\pm$3.0& 70.46$\pm$4.0& 96.62$\pm$1.7& 96.13$\pm$1.2&89.29$\pm$1.7\\ 
MDL-Net  & 57.09$\pm$4   & 61.15$\pm$4.5 &  73.59$\pm$22.4   & 74.92$\pm$2.6& 74.52$\pm$4.8& 70.13$\pm$4.1& 88.16$\pm$2.2& 87.84$\pm$1.9&85.04$\pm$2.5\\ \hline
\textbf{Proposed} &   64.49$\pm$3.9 &   71.55$\pm$4.3   &   78.01$\pm$7.6 & 80.79$\pm$3.3& 75.01$\pm$3.2& 71.76$\pm$4.8& 97.59$\pm$1.2& 97.16$\pm$0.9&90.08$\pm$1.8\\ \hline
\end{tabular}}
\label{tab:comp}
\end{table*}

\begin{table*}[ht]
\centering
\caption{Ablation Studies on the Cricial Designs}
\resizebox{\textwidth}{!}{
\begin{tabular}{c|cccc|c|ccc}
\hline
                         & Temporal Self-Fusion & Spatial Fusion & Modality Refinement&Alignment& M2M Weighting & PR-AUC      & ROC-AUC     & Accuracy   \\ \hline
\multirow{4}{*}{Modules} &                      & $\checkmark$   & $\checkmark$&$\checkmark$& Dot      & 62.09$\pm$7.4 & 66.74$\pm$7.4 & 54.36$\pm$31.8 \\
                         & $\checkmark$         &                & $\checkmark$&$\checkmark$& Dot      & 60.29$\pm$2.7 & 65.36$\pm$1.6 & 63.15$\pm$23.4 \\
                         & $\checkmark$         & $\checkmark$   &             &$\checkmark$& Dot      & 61.76$\pm$5.0 & 64.81$\pm$4.6 & 57.42$\pm$23.1 \\ 
                         & $\checkmark$         & $\checkmark$   &$\checkmark$ &            & Dot      &60.50$\pm$4.7&64.48$\pm$4.7&83.72$\pm$11.2  \\\hline
\multirow{6}{*}{Loss}    & $\checkmark$         & $\checkmark$   & $\checkmark$&$\checkmark$& $\times$ &60.76$\pm$4.2&62.83$\pm$3.4 &80.62$\pm$4.7 \\
                         & $\checkmark$         & $\checkmark$   & $\checkmark$&$\checkmark$& Cosine   & 61.43$\pm$1.8 & 65.33$\pm$2.9 & 72.93$\pm$8.9  \\
                         & $\checkmark$         & $\checkmark$   & $\checkmark$&$\checkmark$& KL       & 60.03$\pm$5.0& 66.15$\pm$5.2 & 58.93$\pm$23.0 \\
                         & $\checkmark$         & $\checkmark$   & $\checkmark$&$\checkmark$& JSD      & 59.50$\pm$3.6 & 64.08$\pm$3.1 & 68.58$\pm$23.6 \\
                         & $\checkmark$         & $\checkmark$   & $\checkmark$&$\checkmark$& MMD      & 60.27$\pm$1.9 & 65.47$\pm$2.4 & 52.00$\pm$28.8 \\
                         & $\checkmark$         & $\checkmark$   & $\checkmark$&$\checkmark$& \textbf{Dot}& 64.49$\pm$3.9 & 71.55$\pm$4.3 & 78.01$\pm$7.6 \\ \hline
\end{tabular}}\label{tab:abl}
\end{table*}

\subsection{Overall Performance}
To demonstrate the superiority of the proposed method, we conduct experiments based on the EHBS dataset and include several commonly used and state-of-the-art 3D multimodal methods for comparison, including:
\begin{itemize}
    \item \textbf{SwiFT-EF}: Direct extension of our backbone by fusing the fMRI and sMRI at the early stage, and using the fused feature as the input for SwiFT.
    \item \textbf{SwiFT-LF}: Direct extension of our backbone by fusing the fMRI and sMRI at the late stage, which combines predictions from each modality encoder.
    \item \textbf{EMV-Net} \cite{wei2023deep}: A model originally proposed for multi-view learning. It is based on CNN-Transfomer, with attention-based cross-modality calibration modules to fuse information.
    \item \textbf{mmFormer} \cite{zhang2022mmformer}: It first encodes features from each modality via CNN, then applies inter-modal Transformers to fuse modalities. A regularizer is included to promote the equivariance between modalities.
    \item \textbf{MDL-Net} \cite{qiu20243d}: A 3D ResNet-based model that fuses modalities in multiple stages, within local and global scales.
\end{itemize}

The results are presented in Table \ref{tab:comp}. Note that here we only use fMRI and sMRI data. As in the table, on the EHBS dataset, the proposed method significantly outperforms others in terms of PR-AUC, except mmFormer. However, the method gains 71.55\% ROC-AUC, much higher than mmFormer. For other baselines, SwiFT-LF achieves reasonable performance compared to others, which shows the superbness of SwiFT as the backbone.

\begin{table*}[htbp]\centering
\caption{Performance on Modality Combinations}
\resizebox{\textwidth}{!}{
\begin{tabular}{ccc|ccc|ccc|ccc}
\multirow{2}{*}{fMRI} & \multirow{2}{*}{sMRI} & \multirow{2}{*}{Tabular} & \multicolumn{3}{c|}{EHBS}                      & \multicolumn{3}{c|}{ADNI}                     & \multicolumn{3}{c}{HCP}                       \\ \cline{4-12} 
                      &                       &                          & PR-AUC        & ROC-AUC       & Accuracy       & PR-AUC        & ROC-AUC       & Accuracy      & PR-AUC        & ROC-AUC       & Accuracy      \\ \hline
$\checkmark$          &                       &                          & 61.12$\pm$4.4 & 64.85$\pm$4.4 & 73.08$\pm$12.5 & 70.128$\pm$4.1     & 68.018$\pm$3.8     & 66.878$\pm$4.2     & 90.168$\pm$1.6     & 91.048$\pm$1.8     & 85.558$\pm$2.1     \\
                      & $\checkmark$          &                          & 62.86$\pm$6.5 & 67.18$\pm$9.0 & 82.53$\pm$2.5  & 78.198$\pm$3.5     & 76.088$\pm$3.3     & 75.828$\pm$4.0     & 96.498$\pm$1.1     & 96.828$\pm$1.5     & 90.018$\pm$2.2     \\
$\checkmark$          & $\checkmark$          &                          & 64.49$\pm$3.9 & 71.55$\pm$4.3 & 78.01$\pm$7.6  & 80.79$\pm$3.3 & 75.01$\pm$3.2 & 71.76$\pm$4.8 & 97.59$\pm$1.2 & 97.16$\pm$0.9 & 90.08$\pm$1.8 \\
$\checkmark$          & $\checkmark$          & $\checkmark$             & 67.66$\pm$5.3 & 73.46$\pm$6.9 & 80.18$\pm$5.7  & 81.088$\pm$2.8     & 77.928$\pm$2.9     & 72.288$\pm$3.3     & -             & -             & -     \\       \hline
\end{tabular}}
\label{tab:mod}
\end{table*}

\begin{figure*}[htpb]
\centerline{\includegraphics[width=\linewidth]{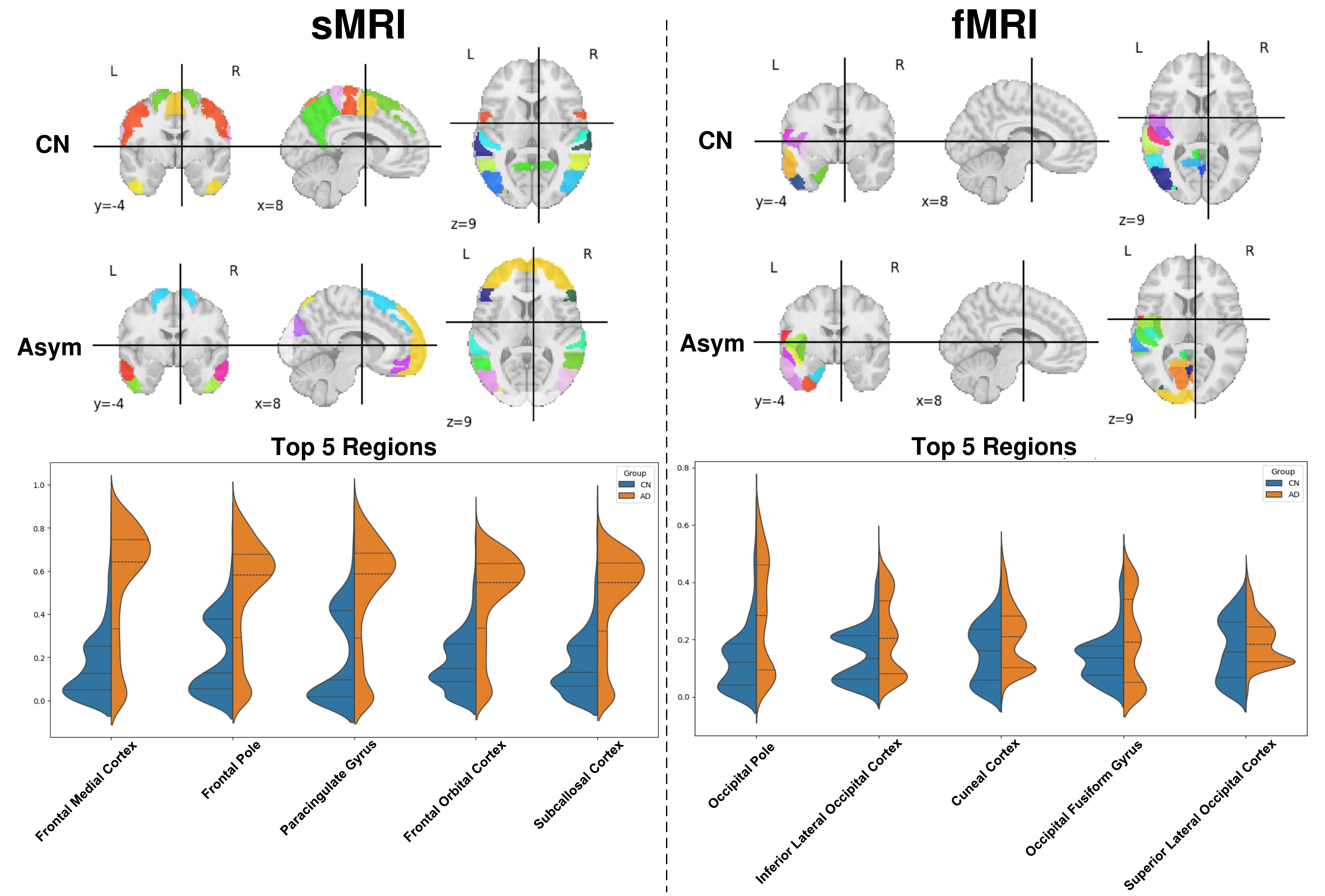}}
\caption{Visualizations of the key brain regions contribute to the framework. We compute the spatial co-attention scores for fMRI and sMRI, map them onto the atlas, then apply a 95\% threshold for better visualization. In the bottom, we show the top five regions with the highest scores. Note that "CN" represents the healthy control, and "AD" represents the patients.}
\label{fig:viz}
\end{figure*}

\subsection{Ablation Study}
We begin by evaluating the impact of different modalities on the performance of the proposed method based on the EHBS dataset. As shown in Table \ref{tab:mod}, on EHBS dataset using only sMRI achieves a PR-AUC of 62.86\% and a ROC-AUC of 67.18\%, which are higher than the results obtained with fMRI alone. Combining both MRI modalities significantly enhances performance, demonstrating the complementary nature of these data sources. Furthermore, incorporating tabular data further improves the PR-AUC from 64.49\% to 67.66\% and the ROC-AUC from 71.55\% to 73.46\%. These results highlight the value of tabular information as a strong biomarker for diagnosing AD. Similar results can be witnessed on other datasets, where the inclusion of tabular data enhances the model's performance, and sMRI plays a more important role than fMRI. Note that HCP does not have tabular data.

To validate our model design, we perform ablation studies on key modules and assess the self-weighting mechanism for the M2M loss based on the EHBS dataset. Specifically, we replace temporal self-fusion with the direct use of raw temporal embeddings, substitute spatial fusion with simple addition, remove the modality refinement modules, and omit the contrastive alignment loss. Additionally, we test how different discrepancy measures for self-weighting affect performance. The results are summarized in  Table \ref{tab:abl}. Note that $\times$ indicates the absence of self-weighting. The fusion and alignment modules are shown to contribute significantly to model performance. In particular, removing spatial fusion or alignment results in the most substantial performance degradation, underscoring their critical roles. For self-weighting, excluding it leads to decreased ROC-AUC and PR-AUC, suggesting that capturing multi-to-multi correspondence between modalities is beneficial. Among different discrepancy measures, cosine similarity yields the best results, whereas JSD performs the worst.

\begin{table}[htbp]\centering
\caption{Model Complexity Analysis}
\begin{tabular}{c|c|c}
Method   & \# Params (M) & FLOPs (G) \\ \hline
SwiFT-EF & 0.52         & 1.11      \\
SwiFT-LF & 1.12         & 32.86     \\
EMV-Net  & 4.72         & 34.37     \\
mmFormer & 35.82        & 86.94     \\
MDL-Net  & 2.87         & 12.56     \\ \hline
Proposed & 1.24         & 34.55    \\ \hline
\end{tabular}\label{tab:complex}
\end{table}

\subsection{Model Complexity}
We also further analyzed the complexity of the model and compared the number of model parameters and floating point operations (FLOPs) between different methods. As in Table \ref{tab:comp}, mmFormer has the largest number of parameters and calculations, making its the best model among the baseline methods (as in Table \ref{tab:complex}). On the other hand, SwinFT-EF is the lightest model as it simply combines the modalities as the input. The proposed method has relatively fewer parameters than other methods, while its FLOPs is higher than SWiFT-EF and MDL-Net. However, our method performs better than these two methods on all three datasets.

\begin{figure}[htpb]
\centerline{\includegraphics[width=\linewidth]{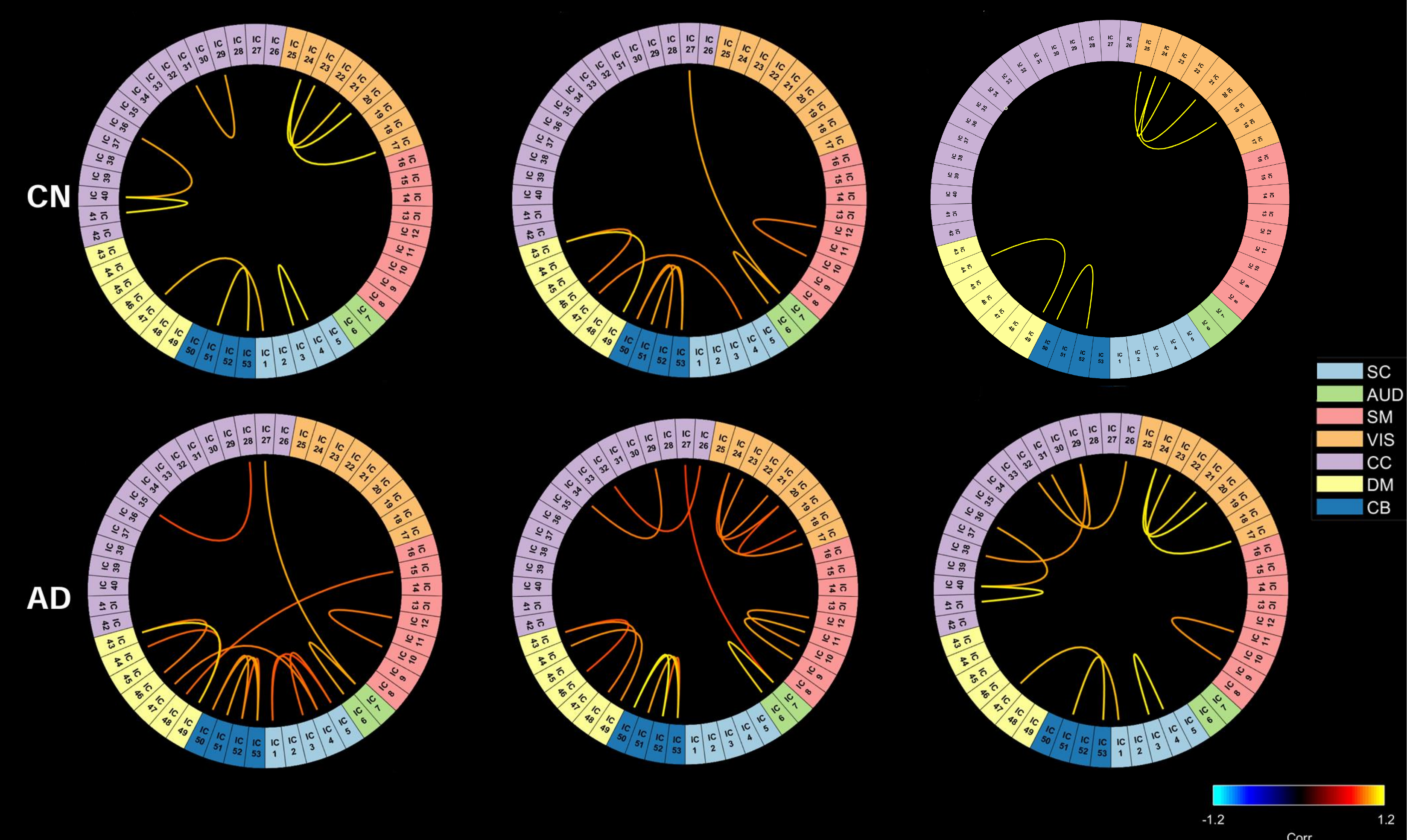}}
\caption{Visualizations of the top 3 brain states contributing to the diagnosis. We compute the temporal latent co-attention scores for fMRI and calculate the functional connectivity for 3 brain states that have the highest scores. Here, "SC" stands for subcortical network,  "AUD" stands for auditory network, "SM" stands for sensorimotor network, "VIS" stands for visual network, "CC" stands for cognitive-control network, "DM" stands for default-mode network, and "CB" stands for cerebellar network.}
\label{fig:fmri}
\end{figure}

\subsection{Discriminative Brain Regions and Connectivities}
To verify the reliability and interpretability of the proposed DRL, we conduct a salient analysis of ROIs obtained by the model using the EHBS dataset. Specifically, we compute the spatial co-attention $\mathcal{H}_{joint}^{sp}$ obtained by Eq. \ref{eq:lFuse} as the complementary features selected from fMRI and sMRI, respectively. The weights are mapped to a standard atlas for better visualization, and we further apply a 95\% threshold for better visualization. We also show the top 5 contributing brain regions for fMRI and sMRI for cognitive normal controls (CN) and Alzheimer's disease patients (AD).

As presented in Fig. \ref{fig:viz}, the  fMRI identified frontal regions (medial cortex, pole, paracingulate/orbital cortex, subcallosal cortex) as key Alzheimer's discriminators. These areas anchor the Default Mode Network (DMN), which shows characteristic connectivity disruptions in Alzheimer's through decreased posterior-anterior synchronization and compensatory frontal hyperactivity. Their prominence aligns with Alzheimer's progression, impairing executive function, emotional regulation, and introspection - processes mediated by frontal DMN hubs \cite{corriveau2024cerebral,wei2024deep}. For sMRI, our model highlighted occipital regions (pole, lateral/fusiform cortices, cuneus) that are fundamentally involved in visual processing. Combined with fMRI, this demonstrates complementary detection of network dysfunction and downstream structural degeneration across Alzheimer's disease stages. These distinct regional profiles underscore the value of the proposed framework, as fMRI and sMRI capture different but complementary pathological signatures of Alzheimer's disease. 

For the temporal co-attention $\mathcal{H}_{f}^{te}$, we provide further analysis to show the brain states that contribute to the framework. Specifically, we compute the top 3 brain states with the highest scores, then compute the correlations between functional regions across the whole brain using a standard fMRI template \cite{du2020neuromark}. As shown in Fig. \ref{fig:fmri}, the intra-correlations in DM, VIS, and CC networks contribute to the diagnosis. Comparing the two groups of subjects, AD patients show excessive focus on CC and DM networks. These observations aligned with our previous findings from Fig. \ref{fig:viz}, suggesting that the temporal co-attention can extract useful functional dynamics from fMRI.

\begin{figure}[htpb]
\centerline{\includegraphics[width=\columnwidth]{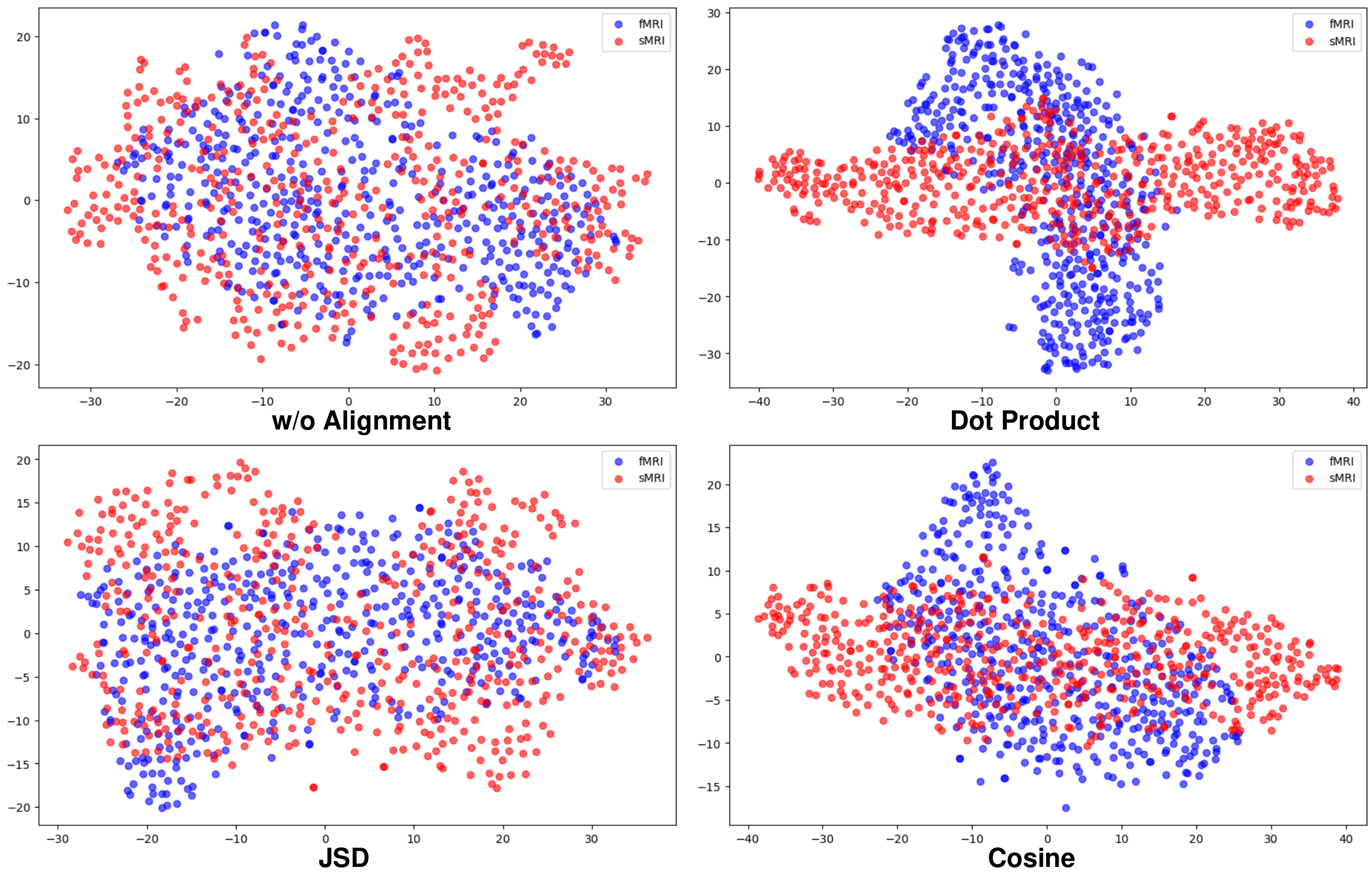}}
\caption{t-SNE visualizations of fMRI and sMRI embeddings in the latent space. The proposed dot-product M2M contrastive alignment produces more concentrated embeddings, with fMRI and sMRI distributions appearing nearly "orthogonal" to each other, indicating effective alignment. In contrast, without alignment or when using JSD for self-weighting, the embeddings show no significant distributional differences.}
\label{fig:tsne}
\end{figure}

\subsection{Visualizing the Aligned Latent Space}
To demonstrate the impact of M2M contrastive alignment, we analyze the latent embeddings of sMRI and fMRI using t-SNE visualizations. Specifically, we compare the embeddings under four conditions: (1) without any alignment, (2) with the proposed dot-product M2M contrastive alignment, (3) using cosine similarity for self-weighting, and (4) using Jensen-Shannon Divergence (JSD) for self-weighting.

As illustrated in Fig. \ref{fig:tsne}, when no alignment is applied, the embeddings from fMRI and sMRI do not exhibit clear distributional differences. A similar pattern is observed when JSD is used for self-weighting, likely because JSD fails to capture the true correspondence between modalities, leading to ineffective alignment. In contrast, both dot-product and cosine-similarity-divergence-based alignments result in more concentrated embeddings for fMRI and sMRI. Notably, these embeddings form distinct distributions with one appearing nearly "orthogonal" to the other, indicating that the modalities retain their unique characteristics while being aligned in a shared space.

\section{Conclusion}
In this work, we propose a novel framework for Alzheimer's Disease (AD) diagnosis that integrates sMRI, fMRI, and tabular data through adaptive multimodal fusion. Our approach leverages co-attention and bottleneck refinement modules to effectively combine complementary information across modalities. To address the inherent heterogeneity among modalities, we introduce a multi-patch-to-multi-patch (M2M) contrastive alignment loss, which aligns patch-wise representations in the latent space, ensuring robust cross-modal correspondence. Comprehensive experiments validate the effectiveness and superiority of our framework, demonstrating improved diagnostic performance. We further show the discriminative brain regions from fMRI and sMRI, aligned with previous clinical research in Alzheimer's. Additionally, t-SNE visualizations provide insights into the impact of our alignment strategy, highlighting its ability to achieve meaningful latent space alignment.

\bibliographystyle{ACM-Reference-Format}
\bibliography{references}


\end{document}